\documentclass{edbk} 
  \usepackage{edbkps}
  \usepackage{graphicx}
\normallatexbib
\begin{document}
\articletitle{Physical perspectives on the global \\
optimization of atomic clusters}
\chaptitlerunninghead{Physical Perspectives on Cluster Optimization}
\author{Jonathan P.\ K.\ Doye}
\affil{University Chemical Laboratory, Lensfield Road, Cambridge CB2 1EW, United Kingdom}
\email{jon@clust.ch.cam.ac.uk}
\begin{abstract}
In this chapter the physical aspects of the global optimization of the geometry 
of atomic clusters are elucidated.
In particular, I examine the structural principles that determine the nature of the 
lowest-energy structure, the physical reasons why some clusters are especially 
difficult to optimize and how the basin-hopping transformation of the potential
energy surface enables these difficult clusters to be optimized.
\end{abstract}
\begin{keywords}
atomic clusters, basin-hopping, multiple funnels
\end{keywords}
\section{Introduction}

Global optimization (GO) is essentially a mathematical task. 
Namely, for the class of GO problems I will be particularly considering here, 
it is to find the absolute minimum (or maximum) of a cost function, $f({\bf x})$, 
where ${\bf x}$ belongs to a subset $D$ of Euclidean n-space,
${\cal R}^n$ \cite{Horst95}, i.e.\
\begin{equation}
\hbox{find }{\bf x^*}\hbox{ such that }f({\bf x^*})\leq f({\bf x}) 
\qquad \forall\,{\bf x}\in D \subset {\cal R}^n .
\end{equation}
Although the applications of global optimization span a wide range of 
fields---from the economics of business in the travelling salesman problem
to biophysics in  the lowest-energy structure of a protein---this 
does not take away from
the essentially
mathematical nature of the optimization problem. 

So why do I wish to discuss the {\it physical} aspects of global optimization? 
To begin with we should realize that even for GO problems that do not correspond 
to a physical system, physical properties can be associated with the system by thinking of the 
cost function as a potential energy function, $E({\rm x})$.
This allows the thermodynamics of the system to be defined. When the system is at equilibrium
at a temperature $T$ each point ${\bf x}$ in configuration space will be sampled with 
a probability proportional to its Boltzmann weight, $\exp(-E({\bf x})/kT)$, where $k$ is the
Boltzmann constant.
Furthermore, for systems with continuous coordinates the forces, $F({\bf x})$, 
associated with each coordinate
can be obtained from the gradient of the cost function, i.e.\ $F({\bf x})=-\nabla E({\bf x})$. 
Once masses are associated with each coordinate, the dynamics are then defined through 
Newton's equations of motion. If one wishes the system's dynamics can then be simulated by 
integrating these equations of motion, as in the molecular dynamics method \cite{FrenkelSmit}.  
Even when the coordinates can only take discrete values, Monte Carlo (MC) simulations can still provide 
a pseudo-dynamics with the number of steps taking the role of time.  

Of course, this connection to physics is most transparent, and most natural, 
when the system being optimized is a physical system, 
which has a real (and potentially observable) thermodynamics and dynamics. 
Furthermore, in those cases where the cost function does truly correspond 
to the potential energy of the system,
there is another physical dimension to the problem---how is the structure of 
the global minimum determined by the physical interactions between the atoms 
and molecules that make up $E({\bf x})$?

Given that we have established that physical properties can be associated with any
system being optimized, what relevance does this physics have to the task of global optimization?
Firstly, many GO algorithm have drawn their inspiration from physics. 
Most famously, simulated annealing is analogous to the slow cooling of a melt to allow the
formation of a near perfect crystal, the idea being that if equilibrium is maintained in the 
simulation as the system is cooled, then at zero temperature it must end up in the global minimum \cite{KirkSA}.
There are many other physically-motivated GO approaches.
The extension of statistical thermodynamics to systems with non-extensive thermodynamics 
through the use of Tsallis statistics \cite{Tsallis88,Tsallis99} has led
to a generalized simulated annealing \cite{Tsallis96,Andricoaei96} 
which is no longer tied to the Boltzmann distribution and 
is often more efficient than standard simulated annealing. 
Genetic algorithms imitate the biophysical evolution of the genome \cite{Goldberg89}.
And I could go on.

However, this is not the link between physics and global optimization 
that is my focus here.
Rather, I wish to show how the ease or difficulty of global 
optimization is often intimately linked to the physics of the system.
The insights obtained from understanding the physical basis for the success or failure 
of an algorithm not only provide an understanding of the limitations of the method
and a basis for assessing the likelihood of success in future applications,
but also aid the development of new algorithms by identifying the main physical 
challenges that need to be overcome to enable greater efficiency and suggesting the 
type of physical behaviour that would need to be incorporated into an improved algorithm.

I will attempt to achieve this aim by concentrating on one class of problems, 
namely the global minimization of the potential energy of an atomic cluster.
Furthermore, I will mainly concentrate on model systems where the cost function is 
computationally cheap to evaluate, 
enabling the physical properties of these systems to be comprehensively examined and understood.
As outlined by Hartke elsewhere in this book \cite{Hartke00}, this class of problems is of great general 
interest to the chemical physics community, because the identification of a cluster's structure is often 
a prerequisite for understanding its other physical and chemical properties.

In this chapter I start at the `end', first showing the structures of the putative global minima\footnote{As 
I cannot prove the optimality of the lowest-energy minima that I find, 
I should refer to the lowest known structures only as {\it putative} global minima, 
but for convenience I usually drop this adjective.}
for a number of cluster systems in order that the reader can understand some of the 
physical principles that determine the structure and 
how these relate to the interatomic interactions. 
Furthermore, the structure provides a basis for understanding a cluster's 
thermodynamic and dynamic properties,
especially when, as in some of our examples, the competition between different 
structural types plays an important role.

I then consider some of the GO algorithms that are most successful for these systems focussing on
those that use the basin-hopping transformation of $E({\bf x})$ \cite{WalesD97} and
on how the performance of these algorithms depend on the system and the cluster size.
I then look at the physical properties of some of the clusters, 
relating these back to the ease or difficulty of global optimization. 
I firstly examine the topography of the multi-dimensional surface defined by $E({\bf x})$ 
(the so-called potential energy surface (PES) or energy landscape), then the thermodynamics and dynamics.
Finally, I show why basin-hopping is able to locate the global minimum 
in those clusters where the PES has a multiple-funnel topography, 
and make some suggestions as to how further gains in efficiency might be secured.

\section{Cluster structure}

In this section I mainly concentrate on the structures of model clusters, 
where the interactions have simple idealized forms that are isotropic, thus favouring compact
geometries. The models have been chosen so that they span a wide range of structural behaviour
that is likely to be relevant to rare gas, metal and molecular clusters bound by dispersion 
forces, but not to clusters with directional covalent bonding or molecular clusters with 
directional intermolecular forces, as with the hydrogen bonding in water clusters.

\begin{figure}
\begin{center}
\includegraphics[width=9cm]{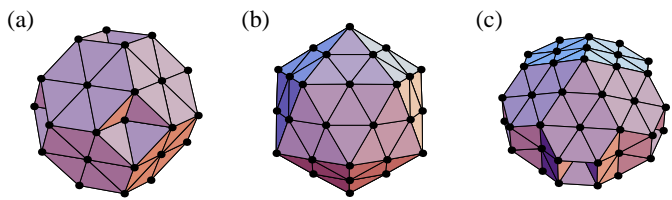}
\end{center}
\vglue -0.85cm
\caption{\label{fig:structure_main}
Three examples of the structures clusters can adopt: (a) a 38-atom truncated octahedron,
(b) a 55-atom Mackay icosahedron, and (c) a 75-atom Marks' decahedron. 
These clusters have the optimal shape for the three main types of regular packing seen in clusters:
close-packed,
icosahedral and decahedral, respectively.
The Mackay icosahedron is a common structure and is observed for rare gas \cite{Harris84} 
and many metal \cite{Martin96} clusters.
The truncated octahedron has been recently observed for nickel \cite{Parks97} and gold \cite{Alvarez97}
clusters, and the Marks decahedron for gold clusters \cite{Cleveland98}.}
\vglue -0.3cm
\end{figure}

Most of the clusters we consider have only pair interactions, i.e.\ 
\begin{equation}
E({\bf x})=\sum_{i<j} V(r_{ij}),
\end{equation}
where $V$ is the pair potential and $r_{ij}$ is the distance between atoms $i$ and $j$.
In this case we can partition the energy into three terms \cite{Doye95c}:
\begin{equation}
\label{eq:Esplit}
E=-n_{\rm nn}\epsilon+E_{\rm strain}+E_{\rm nnn}
\end{equation}
where $\epsilon$ is the pair well depth, $n_{\rm nn}$ is the number of nearest neighbours,
\begin{equation}
E_{\rm strain}=\sum_{i<j,r_{ij}<r_0} (V(r_{ij})-V(r_{\rm eq}))
,\qquad
E_{\rm nnn}=\sum_{i<j,r_{ij}\ge r_0} V(r_{ij}),
\end{equation}
and $r_{\rm eq}$ is the equilibrium pair distance. 
Two atoms are defined as nearest neighbours if $r_{ij}<r_0$. $r_0$ should lie between the first and second 
coordianation shells and a typical value would be $1.35\, r_{\rm eq}$, although the exact value is somewhat arbitrary.
The first term in Equation \ref{eq:Esplit} is the ideal pair energy if all 
$n_{\rm nn}$ nearest-neighbour pairs lie exactly at the equilibrium pair distance,
the strain energy is the energetic penalty for the deviation of nearest-neighbour distances
from the equilibrium pair distance and $E_{\rm nnn}$ is the contribution to the energy from 
non-nearest neighbours. 

$E_{\rm nnn}$ is usually smaller than the other two terms and is relatively independent
of the detailed structure. 
Therefore, the global minimum usually represents the best balance between maximizing $n_{\rm nn}$ 
and minimizing $E_{\rm strain}$. 
For an atom in the interior of a cluster this is usually achieved through the atom having 
a coordination number of twelve. 
This can be achieved as in close-packing, but another possibility is
an icosahedral coordination shell. $n_{\rm nn}$ is further increased through the cluster having a
compact spherical shape, and through the surface mainly consisting of faces with a 
high co-ordination number. For example, an atom on a face-centred-cubic (fcc) $\{111\}$ face
has nine nearest neighbours, whereas an atom on a $\{100\}$ face has eight nearest neighbours.\footnote{
The fcc $\{111\}$ planes corresponds to the close-packed planes, where the atoms in the planes form a grid of 
equilateral triangles. The atoms in the $\{100\}$ planes form a square grid.}

\begin{figure}
\begin{center}
\includegraphics[width=7cm]{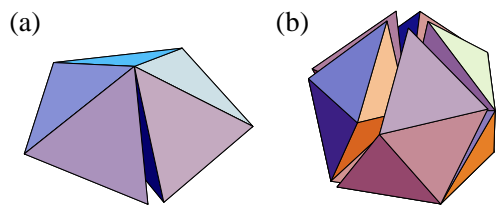}
\end{center}
\vglue -1.0cm
\caption{\label{fig:gaps}
Examples of the strain involved in packing tetrahedra. (a) Five regular tetrahedra
around a common edge produce a gap of $7.36^\circ$. (b) Twenty regular tetrahedra about a common vertex
produce gaps equivalent to a solid angle of 1.54 steradians.}
\vglue -0.3cm
\end{figure}

The three main types of cluster structure found for systems with isotropic interactions, namely
icosahedral, decahedral and close-packed\footnote{I use the term 
{\em close-packed} to refer to any structure where all the interior atoms
of the cluster have a face-centred-cubic or hexagonal close-packed coordination shell. 
This definition allows for any stacking sequence of close-packed planes, 
but does not admit any configuration of twin planes that must involve strain.}
structures, are depicted in Figure \ref{fig:structure_main}. 
These examples have the optimal shape for each structural type, and all have been identified
experimentally.

One of the unusual properties of clusters is that they can exhibit non-crystallographic symmetries,
because there is no requirement for translational periodicity.
Decahedral clusters have a single five-fold axis and are based on a pentagonal bipyramid that
can be thought of as five strained fcc tetrahedra sharing a common edge. 
The symmetry axis corresponds to this common edge.
The Marks decahedron \cite{Marks84}, which represents the optimal shape for this structural type,
can be formed from a pentagonal bipyramid by exposing $\{100\}$ faces at the equatorial edges 
then introducing reentrant $\{111\}$ faces.
Mackay icosahedra \cite{Mackay} have six five-fold axes of symmetry and can be thought of 
as twenty strained fcc tetrahedra sharing a common edge.
The fcc cluster represented in Figure \ref{fig:structure_main}a is 
simply a fragment of the bulk fcc lattice. 

Icosahedral structures generally have the largest $n_{\rm nn}$ because of their spherical shape
and $\{111\}$ faces, and close-packed clusters the smallest $n_{\rm nn}$ because of their 
higher proportion of $\{100\}$ faces. By contrast, close-packed clusters can be unstrained,
whereas, as Figure \ref{fig:gaps} illustrates, decahedra and icosahedra are increasingly strained. 
The strain energy is proportional to the volume of the cluster, but differences in $n_{\rm nn}$ 
are due to surface effects. Therefore, icosahedra are likely to be found at small sizes, 
but at sufficiently large size, the cluster must take on the bulk structure.
At intermediate sizes decahedra can be most stable. The sizes at which the crossovers between 
structural types occur is system dependent.

\begin{figure}
\begin{center}
\includegraphics[width=11cm]{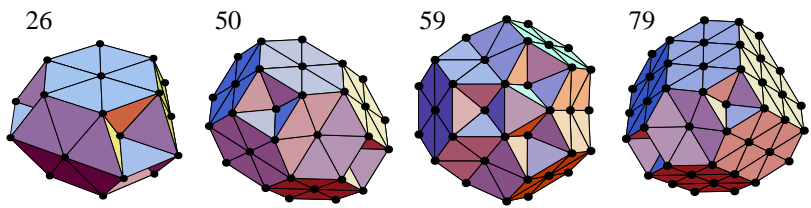}
\end{center}
\vglue -0.8cm
\caption{\label{fig:cp} Four examples of stable close-packed structures for $N<80$.
The sizes are as labelled. The 59-atoms structure has $T_d$ point group symmetry, and the
rest $D_{3h}$.}
\vglue -0.3cm
\end{figure}

The structures illustrated in Figure \ref{fig:structure_main} 
involve exactly the right number of atoms to form a cluster of the optimal shape. 
For example, complete Mackay icosahedra can be formed at $N=13, 55, 147, 309, \dots$ 
At intermediate sizes clusters have an incomplete surface layer.
Marks decahedra with square $\{100\}$ faces occur at $N=75, 192, 389, \dots$. 
Other complete Marks decahedra that are less spherical can be found in between 
these sizes, e.g.\ at $N$=101 and 146. 
Fcc truncated octahedra with regular hexagonal $\{111\}$ faces can be found at 
$N=38, 201, 586 \dots$, and other less spherical truncated octahedra can be found, for
example, at $N$=79, 116, 140 \cite{Doye95d}. 
Furthermore, because the energy of a twin plane is often small, close-packed structures 
with other forms can also be particularly stable. Four examples 
are given in Figure \ref{fig:cp} for $N<100$. 
The 26-atom structure has a hexagonal close-packed (hcp) structure;
the 50-atom structure consists of two fragments of the 38-atom structure joined at a twin plane;
the 59-atom structure consists of a 31-atom fcc truncated tetrahedron with each face covered by 
a 7-atom hexagonal overlayer that occupies the hcp surface sites; and 
the 79-atom structure, which is similar to the 50-atom structure, is formed by the introduction
of a twin plane into the 79-atom truncated octahedron \cite{Raoult89a}. 

\begin{figure}
\begin{center}
\includegraphics[width=6cm]{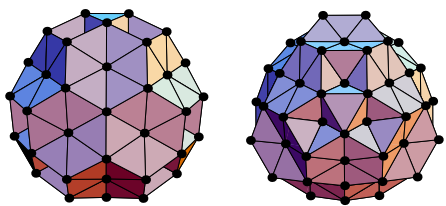}
\end{center}
\vglue -0.9cm
\caption{\label{fig:Leary_Td}
Front and back views of the 98-atom Leary tetrahedron.}
\vglue -0.3cm
\end{figure}

Recently, a new structural type called a Leary tetrahedron \cite{Leary99} has been discovered. 
An example with 98 atoms is illustrated in Figure \ref{fig:Leary_Td}.
At the centre of this structure is an fcc tetrahedron. 
To each of the faces of this tetrahedron, further fcc tetrahedra (minus an apical atom) 
are added, to form a stellated tetrahedron.
Finally, the edges of the original tetrahedron are covered by 7-atom hexagonal overlayers.
The coordination along the edges of the central tetrahedron is the same as along the
symmetry axis of the decahedron and so the strain energy of this structure is intermediate 
between icosahedra and decahedra.
It is not yet clear how general this class of structures is. 
The 98-atom example is the global minimum for a model potential \cite{Leary99} 
and mass spectroscopic studies of clusters of C$_{60}$ molecules
suggest that (C$_{60}$)$_{98}$ has this structure \cite{Branz00}. However, it may
be that the stability of this structural class is restricted to $N$=98, because this size results in
a particularly spherical shape, and that equivalent structures at larger sizes 
(e.g.\ $N$=159, 195) are never competitive.

\subsection{Lennard-Jones clusters}

In this section I focus on clusters bound by the Lennard-Jones (LJ) potential \cite{LJ}:
\begin{equation}
E = 4\epsilon \sum_{i<j}\left[ \left(\sigma\over r_{ij}\right)^{12} - \left
(\sigma\over r_{ij}\right)^{6}\right],
\end{equation}
where $\epsilon$ is the pair well depth and $2^{1/6}\sigma$ is the equilibrium pair separation.
The potential is illustrated in Figure \ref{fig:potentials}a,
and provides a reasonable description of the 
interatomic interactions of rare gases, such as argon.
LJ clusters have become probably the most common test system for GO algorithms
for configurational problems. The number of papers with applications to this model system
is now very large, but unfortunately many are distinctly unimpressive, only reporting 
results for small sizes or failing for relatively simple cases. I do not attempt to review
this literature, but instead refer the interested reader elsewhere \cite{Wille00}.

\begin{figure}
\begin{center}
\includegraphics[width=9cm]{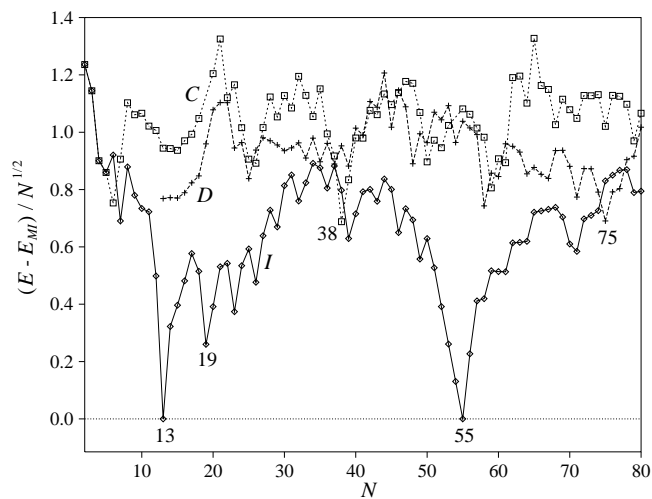}
\end{center}
\vglue -0.5cm
\caption{\label{fig:LJ_EvN} Comparison of the energies of icosahedral (I), decahedral (D) 
and close-packed (C) LJ$_N$ clusters.
The energy zero is $E_{\rm MI}$, a function fitted to the energies of
the first four Mackay icosahedra at $N$=13, 55, 147 and 309:
}
\vglue -0.3cm
\end{figure}

At small sizes the LJ potential is able to accommodate the strain associated 
with icosahedral packing relatively easily \cite{Northby87}. Indeed, only for $N>1600$ are the majority 
of global minimum expected to be decahedral and the crossover to fcc clusters has been 
estimated to occur at $N\approx 10^5$ \cite{Raoult89a}.
This preference for icosahedral packing is also evident from
Figure \ref{fig:LJ_EvN} where I compare the energies of icosahedral, decahedral 
and close-packed clusters. Sizes where complete Mackay icosahedra are possible ($N$=13, 55)
stand out as particularly stable. 
The icosahedra are least stable when the overlayer is roughly half-filled.
Therefore, when especially stable non-icosahedral clusters coincide with these sizes 
there is a possibility that the global minimum will be non-icosahedral.
There are eight such cases for $N\le 147$. 
At $N$=38 the global minimum is the fcc truncated octahedron \cite{Doye95c,Gomez94,Pillardy};
at $N$=75--77 \cite{Doye95c} and 102--104 \cite{Doye95d} the global minima are Marks decahedra;
and at $N$=98 the global minimum is a Leary tetrahedron \cite{Leary99}. 
At these sizes the lines for the decahedral or close-packed 
structures in Figure \ref{fig:LJ_EvN} dip just below the line for the icosahedra.
For $148\le N \le 309$ there are a further eight non-icosahedral global minima \cite{Hartke00,Romero99}, 
all of which are decahedral and which divide into two sets that are based on the complete Marks decahedra 
possible at N=192 and 238.

\subsection{Morse clusters}

\begin{figure}
\begin{center}
\includegraphics[width=11cm]{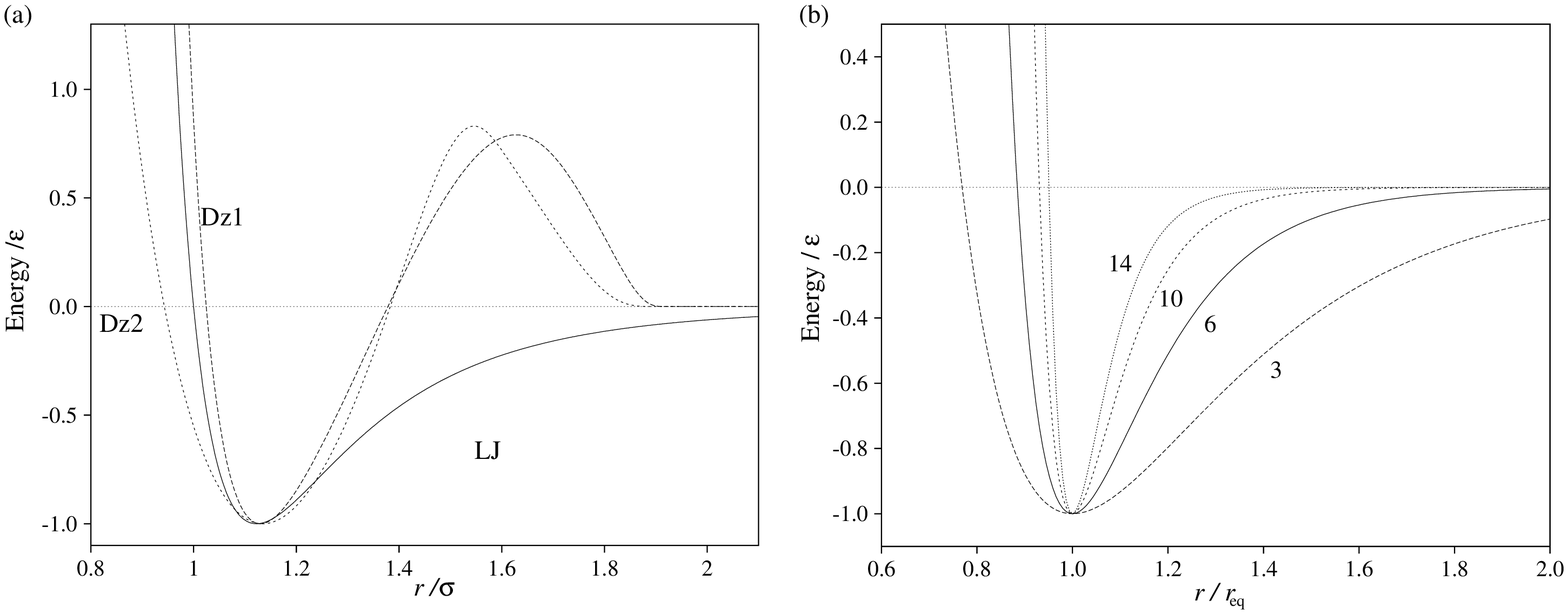}
\end{center}
\vglue -0.5cm
\caption{\label{fig:potentials} (a) A comparison of the Lennard-Jones (LJ) potential with
the Dzugutov potential (Dz1) and a modified version of it (Dz2).
(b) The Morse potential for several values of the range parameter, $\rho$.
}
\vglue -0.3cm
\end{figure}

In this section I focus on clusters bound by the the Morse potential \cite{Morse}: 
\begin{equation}
V_M = \epsilon\sum_{i<j} e^{\rho(1-r_{ij}/r_{\rm eq})}(e^{\rho(1-r_{ij}/r_{\rm eq})}-2),
\end{equation}
where $\epsilon$ is the pair well depth and $r_{\rm eq}$ is the equilibrium pair separation.
In reduced units there is a single adjustable parameter, $\rho$, which determines the range of 
the interparticle forces.
Figure \ref{fig:potentials}b shows that decreasing $\rho$ increases the range of the
attractive part of the potential and softens the repulsive wall, thus widening the potential well.
Values of $\rho$ appropriate to a range of materials have been catalogued elsewhere \cite{WalesMD96}.
The LJ potential has the same curvature at the bottom of the well as the Morse potential when $\rho$=6.
Girifalco has obtained an intermolecular potential for C$_{60}$ molecules \cite{Girifalco} that
is isotropic and short-ranged relative to the equilibrium pair separation
with an effective value of $\rho$=13.62 \cite{WalesU94}.
The alkali metals have longer-ranged interactions, for example
$\rho$=3.15 has been suggested for sodium \cite{GirifalcoW}.

\begin{figure}
\begin{center}
\includegraphics[width=11.5cm]{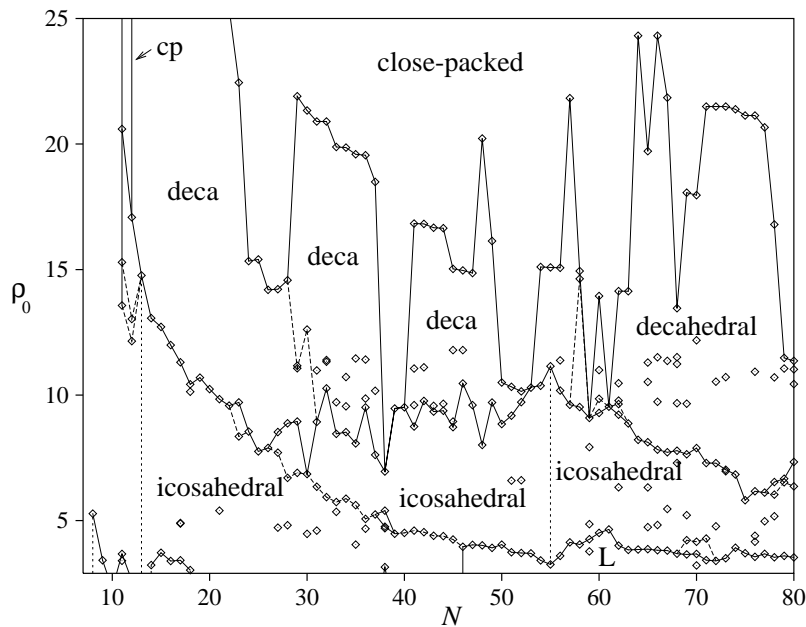}
\end{center}
\vglue -0.6cm
\caption{\label{fig:Mphased}Zero temperature `phase diagram' showing the variation of the lowest-energy
structure with $N$ and $\rho$.
The data points are the values of $\rho$ at which the global minimum changes.
The lines joining the data points divide the phase diagram into regions where the global minima
have similar structures.
The solid lines denote the boundaries between the four main structural types---icosahedral,
decahedral, close-packed and those associated with low $\rho$ (L)---and the dashed lines are internal
boundaries within a structural type, e.g.~between icosahedra with different overlayers, 
or between decahedra with different length decahedral axes.}
\vglue -0.3cm
\end{figure}

The global minima for this system have been found as a function of $\rho$ for all sizes up to $N$=80 
\cite{Doye95c,Doye97d}.\footnote{One of the values reported in Table 1 of Ref.\ \cite{Doye97d} has 
been superseded. At $N$=30 and $\rho$=14 the energy of the global minimum is -106.835\,790.}
Equation (\ref{eq:Esplit}) enables us to understand the effect of $\rho$ on cluster structure.
As $\rho$ increases and the potential well narrows, the energetic penalty for distances deviating
from the equilibrium pair separation increases. 
Thus, $E_{\rm strain}$ increases for strained structures, and so icosahedral and decahedral structures become
disfavoured as $\rho$ increases. This is illustrated in Figure \ref{fig:Mphased}, which shows
how the structure of the global minimum depends on $N$ and $\rho$. The global minimum generally change from 
icosahedral to decahedral to close-packed as $\rho$ is increased. It can be seen that the value of $\rho$
appropriate for the LJ potential lies roughly in the middle of the icosahedral region of Figure \ref{fig:Mphased}.

Alternatively, the effect of $\rho$ can be thought of in terms of its effect on the crossover sizes at 
which a particular structural type becomes dominant. As $\rho$ increases, 
the less strained structures become dominant at smaller sizes.
These effects can also be found in real materials. For example, sodium clusters have been shown to 
exhibit icosahedral structures up to at least 22\,000 atoms \cite{Martin90}, whereas the thermodynamically stable structure of 
clusters of C$_{60}$ molecules have recently been shown to be non-icosahedral for $N>30$ \cite{Branz00}. 

As well as these trends, Figure \ref{fig:Mphased}, of course, also reflects the
specifics of the structures that are possible at each size, so the boundaries between structural types
are not smooth lines but show a lot of detailed structure. For example, the range of $\rho$ values for which
icosahedral structures are most stable is a local maximum at $N$=55 because of the complete Mackay 
icosahedra possible at this size. 
At sizes where close-packed structures have a greater or equal number of nearest neighbours to the 
best decahedral structure, the global minimum changes directly from icosahedral to close-packed.

A new class of structures appears in the bottom right-hand corner of Figure \ref{fig:Mphased}.
They are polytetrahedral clusters with disclination lines running through them.
A polytetrahedral structure can be decomposed into tetrahedra without any interstices.
The 13-atom icosahedron is an example, and one of the possible ways of adding atoms to the 
surface of the icosahedron---the so-called anti-Mackay overlayer---continues the polytetrahedral 
packing. This overlayer does not lead to the next Mackay icosahedron but instead to the
45-atom rhombic tricontahedron (Figure \ref{fig:ptet}a), which can be thought of as an 
icosahedron of interpenetrating icosahedra.
If one imagines adding regular tetrahedra to the form in Figure \ref{fig:gaps}b
one soon realizes that the 45-atom structure must be extremely strained, and 
for this reason it is the global minimum only at low $\rho$,
where this strain can be accommodated. 
For $N>45$ similar polytetrahedral clusters can be formed but based not on the 13-atom icosahedron
but on polyhedra with a higher coordination number. The two examples in Figure \ref{fig:ptet}a
have a 14- and 16-coordinate central atom. These structures can be described in 
terms of disclination lines, where the lines pass along those nearest-neighbour contacts that are
the common edge for six tetrahedra \cite{NelsonS}. 
These types of structures might be thought to be fairly esoteric, but 
they form the basis for the crystalline Frank-Kasper phases \cite{FrankK58,FrankK59} 
where atoms of different size create a preference for coordination numbers higher than 12, 
and so they might be good candidate structures for certain mixed metal clusters. 
Furthermore, they have recently been observed in a model of clusters of heavy metal atoms \cite{Cune00},
and recent experimental diffraction and electron microscopy data for small cobalt clusters can 
best be modelled by a disclinated polytetrahedral structure that is a fragment of a Frank-Kasper
phase \cite{Dassenoy00}. However, at the size corresponding to this experiment ($N\approx 150$)
the long-ranged Morse clusters have disordered polytetrahedral global minima.

\subsection{Dzugutov clusters}

In contrast to the potentials that we have so far examined, the Dzugutov potential 
\cite{Dzugutov91,Dzugutov93b} has a maximum 
that penalizes distances near to $\sqrt 2$ times the equilibrium pair distance 
(Figure \ref{fig:potentials}a), 
the distance across the diagonal of the octahedra in close-packed structures. 
This maximum loosely resembles the first of the Friedel oscillations \cite{Pettifor} often found in 
effective metal potentials.
The potential was originally designed to suppress crystallization in bulk simulations so that
the properties of supercooled liquids and glasses could more easily be studied 
in a one-component system \cite{Dzugutov91,Dzugutov92}. 
However, under certain conditions it was found that a
dodecagonal quasicrystal could be formed on freezing \cite{Dzugutov93}.

\begin{figure}
\begin{center}
\includegraphics[width=10cm]{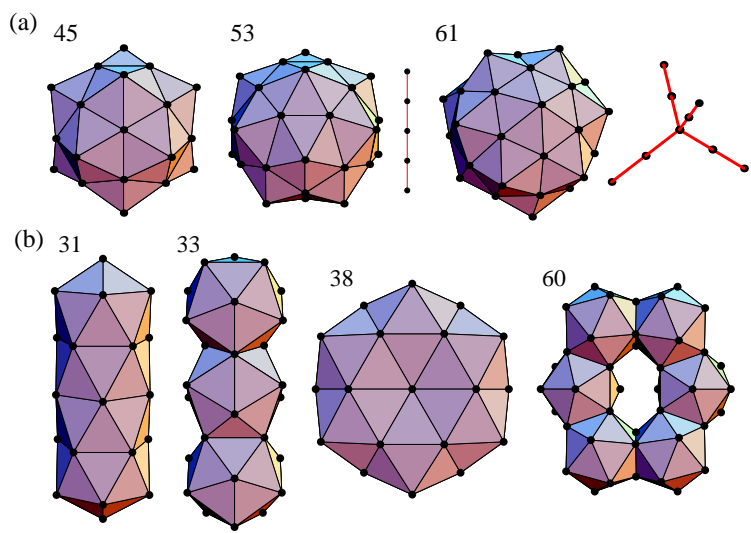}
\end{center}
\vglue -0.5cm
\caption{\label{fig:ptet} Polytetrahedral structures that are the global minima 
for the (a) long-ranged Morse potential and (b) Dzugutov potential. 
Sizes as labelled. The point group symmetries of the structures in (a) 
are $I_h$, $D_{6d}$ and $T_d$ and in (b) are $D_{5h}$, $D_{3d}$, $D_{6h}$ and $C_s$. 
The disclination networks associated with the 53- and 61-atom structures are illustrated
next to the clusters.
}
\vglue -0.3cm
\end{figure}

For clusters the potential will penalize close-packed, decahedral and Mackay icosahedral
structures (the latter two because octahedra are found within the fcc tetrahedral units 
from which the structures are made) and will favour polytetrahedral clusters.
Therefore, one might think that this potential would provide a good model for small cobalt clusters. 
However, as can be seen from Figure \ref{fig:potentials}a the potential is narrower than 
the LJ potential, and matching the second derivative at the equilibrium pair separation
to that of the Morse potential gives an effective value of $\rho$ of 7.52. Therefore,
the potential cannot accommodate the strain in compact polytetrahedral clusters. Instead,
the global minima are non-compact polytetrahedral structures, such as the needles, disc and 
torus illustrated in Figure \ref{fig:ptet}b \cite{Doye01a}. 
These structures are made up of face-sharing or interpenetrating 13-atom icosahedral units. 
In terms of Equation (\ref{eq:Esplit}) they represent the best balance between maximizing $n_{\rm nn}$ 
whilst minimizing both $E_{\rm strain}$ and $E_{\rm nnn}$, where the latter now corresponds to the total 
energetic penalty for distances close to $\sqrt{2}\,r_{\rm eq}$.

In order to generate a model that exhibits ordered compact polytetrahedral
clusters a modified Dzugutov potential was constructed with an effective value of $\rho$ of 5.16,
allowing it to accommodate more strain (Figure \ref{fig:potentials}a).
Preliminary results indicate that the global minima do have the desired structural type, 
and so this system should be useful in generating realistic candidate structures 
to compare with the cobalt experiments \cite{Doye00f}.

\subsection{Comparison with experiment}

One of the remarkable features of these simple model potentials is that 
the structures they exhibit do provide good candidates for the 
structures of real clusters. Indeed, frequently the structures first identified for
these model systems are subsequently identified in experiments. 
For example, the special stability of the structures exhibited by 
the fcc Ni$_{38}$ \cite{Parks97} and Au$_{38}$ \cite{Alvarez97},
the decahedral Au$_{75}$ \cite{Cleveland98} and the tetrahedral (C$_{60}$)$_{98}$ \cite{Branz00} were 
first identified through calculations on LJ and Morse clusters \cite{Doye95c,Leary99}.

Furthermore, as experiments can rarely identify a cluster's structure directly, but often have 
to rely on comparison with properties calculated using candidate structures,
it is extremely useful to have databases of plausible structures available. 
This is the philosophy behind internet repositories such at the Cambridge Cluster Database
(http://brian.ch.cam.ac.uk/CCD.html), which contains the global minima for all the 
potentials described here, and the Birmingham Cluster website (http://www.tc.bham.ac.uk/bcweb/).

In comparisons between experiment and theory the role of temperature and kinetics should 
be remembered. The global minimum is only rigorously the equilibrium structure at
zero temperature. At higher temperatures other structures may become more stable
due to entropic effects \cite{Doye98a,Doye00d} as we will see in Section \ref{sect:therm}. 
Furthermore, it is not always clear whether equilibrium has been 
achieved under the experimental conditions, especially for clusters formed at low temperature 
\cite{Branz00,Baletto00}. 

\section{Global Optimization approaches}
\label{sect:GO}

The type of GO algorithms in which I am interested are those that find global minima, 
not those that are also able to {\em prove} that the best structure found is in fact truly 
global.
Unsurprisingly, the latter is a much more demanding task. For example, 
for LJ clusters good putative global minima have been found up to $N$=309,
but only up to $N$=7 have these structures been proven to be global \cite{Maranas92}.
Of course, the problem with settling for obtaining putative global minima is that
it is difficult to know when to give up looking for a lower-energy solution.
For example, to my surprise, at least, a new putative global minimum was 
recently found for LJ$_{98}$ \cite{Leary99}, even though powerful GO algorithms had 
previously been applied to this cluster \cite{WalesD97,Deaven96,Hartke99,Wolf98}. 
The failure of these previous attempts to locate the global minimum was not because the 
algorithms are unable to locate the Leary tetrahedron, but simply because the computations 
had been terminated too soon. 

I also wish to concentrate on GO algorithms that are unbiased, i.e.\ those that 
do not artificially bias the system towards those structures that physical insight would 
suggest are low in energy, for example, by seeding the algorithm with fragments of 
a certain structural type \cite{Wolf98} or by searching on a lattice for a 
specified structural type \cite{Northby87}. 
Of course, such biased algorithms are usually more efficient.
For example, most of the LJ global minima were first found using such methods \cite{Northby87,Romero99}. 
However, they cannot cope with `surprising' structures that fall outside the expected categories,
and lack transferability to other systems, because they have sacrificed
generality for greater efficiency in the specific problem instances. 
Furthermore, they require a sufficient prior understanding of the structure. 
This may be possible for the model potentials we consider here, but it is a much more 
difficult task with the complex interactions that are often necessary 
to realistically describe a system.

Virtually all the global optimization algorithms that are most successful at 
locating the global minima of clusters have a common feature. Namely, they
make extensive use of local minimization. All GO algorithms 
require elements of both local 
and global 
search. The algorithm has both to be able to explore all regions of configuration space 
(overcoming any energy barriers that might hinder this)
whilst also sufficiently sampling the low-energy configurations within each region.
Performing local minimizations from configurations generated by a global search 
is one way of combining these two elements.

Simulated annealing provides a perhaps more traditional way of achieving this goal.
In simulated annealing, by varying a parameter, the temperature, the nature of the search
is changed from global (at high temperature) to local (as $T\rightarrow 0$). 
However, this approach has a number of weaknesses. 
There is effectively only one local minimization, so if the configuration
does not become confined to the basin of attraction of the global minimum as the temperature is reduced
the algorithm will fail, even if the system had passed through that basin of attraction at higher temperature.
This condition for success is unnecessarily restrictive and leads to inefficiency. 

\begin{figure}
\begin{center}
\includegraphics[width=7cm]{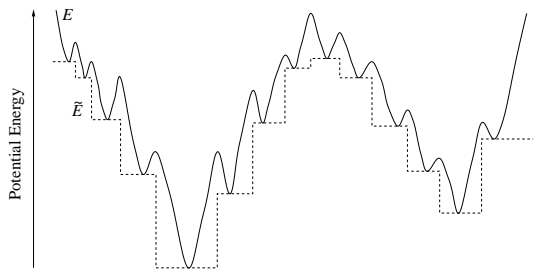}
\end{center}
\vglue -0.5cm
\caption{\label{fig:trans}
A schematic diagram illustrating the effects of the basin-hopping potential
energy transformation for a one-dimensional example.
The solid line is the potential energy of the original surface and the dashed line is the
transformed energy, $\tilde E$.}
\vglue -0.3cm
\end{figure}

There is a further element to the most successful algorithms, namely that the energies of the 
local minima, not of the configurations prior to minimization, are the basis for comparing and
selecting structures. This approach was first used in 1987 by Li and Scheraga in the
application of their `Monte Carlo plus minimization' to polypeptides \cite{Li87a}.
However, despite this approach being independently adopted a number of times subsequently \cite{Xue94b,Deaven95},
it was only in 1997 that it was realized that in this approach one is effectively searching
a transformed PES, $\tilde{E}({\bf x})$, where the energy associated with
each point in configuration space is that of the minimum obtained by a local minimization from that point \cite{WalesD97},
i.e.\ 
\begin{equation}
\tilde{E}({\bf x})={\rm min}\{E({\bf x})\},
\end{equation}
where `min' signifies that an energy minimization is carried out starting from ${\bf x}$.
Unlike many PES transformations proposed in the name of global optimization, 
this `basin-hopping' transformation is guaranteed to preserve the identity of the global minimum.
The transformation maps the PES onto a set of interpenetrating staircases 
with plateaus corresponding to the basins of attraction of each minimum 
(i.e.\ the set of configurations which lead to a given minimum after optimization).
A schematic view of the staircase topography that results from this transformation
is given in Figure \ref{fig:trans}.

The potential advantages of using the basin-hopping transformation 
become clear when we contrast the 
inter-minimum dynamics on the original and transformed PESs. 
In molecular dynamics simulations on the original PES much time is wasted as
the system oscillates back and forth within the well surrounding a minimum, waiting
for the kinetic energy to become sufficiently localized along the direction of a
transition state valley to enable the system to pass into an adjacent minimum.
A similar dynamical (albeit as a function of steps rather than time) picture holds for MC when only local moves
are used. The biased random walk is confined to the well around a minimum, 
frequently being reflected back off the walls of this well,
until by chance the system happens to wander over a transition valley into a new minimum.
However, a completely different picture is appropriate to the dynamics 
in simulations (using MC or discontinuous molecular dynamics\footnote{The 
forces are zero except when the system reaches the edge of a basin of attraction 
at which point the cluster receives an impulse.
Such force fields can be handled in discontinuous molecular dynamics \cite{Alder59}), a method that 
has often been used in dynamical simulations of systems of hard bodies.} \cite{Alder59})
on the transformed PES.
The transformation removes vibrational motion (the Hessian has no positive eigenvalues)
and transitions out of a basin are possible anywhere along the boundary of the basin. 
Therefore, steps in any direction can lead directly to a new minimum. 
Furthermore, downhill transitions are now barrierless. However, as Figure \ref{fig:trans} illustrates, 
significant barriers between low-energy minima can remain if they are separated by high-energy 
intervening minima.

Consequently, on $\tilde{E}({\bf x})$ the system can hop directly between basins; 
hence the name of this transformation. 
Furthermore, much larger MC steps can be taken on $\tilde{E}({\bf x})$; 
such steps would virtually always be rejected on 
the original PES because
atoms would become too close and an extremely high energy would result.
After the transformation atoms can even pass through each other. 

The method of searching $\tilde{E}({\bf x})$ is of secondary importance compared
to the use of the transformation itself. Indeed, the performance is fairly similar for 
the two main methods used, genetic algorithms and the constant temperature MC used in
basin-hopping. Here, we mainly concentrate on the basin-hopping approach and 
refer readers to Hartke's chapter for more detail on the genetic algorithm methodology \cite{Hartke00}.

In the basin-hopping or Monte Carlo plus minimization method, standard Metropolis MC is used, 
i.e.\ moves are generated by randomly perturbing the coordinates, and are always accepted if
$\tilde{E}$ decreases and are accepted with a probability $\exp(-\Delta\tilde{E}/kT)$ if $\tilde{E}$ increases. 
Using constant temperature is sufficient, since there is no great advantage to using an
annealing schedule because the aim is not to trap the system in the global minimum,
but just to visit it at some point in the simulation. One of the advantages of this method
is its simplicity---there are few parameters to adjust. 
It is usually satisfactory to dynamically adjust the step size to produce a 50\% acceptance ratio. 
An appropriate temperature also needs to be chosen, but fortunately the temperature window
for which the method is effective is usually large, and can be quickly found after some experimentation.
Furthermore, there is a well-defined thermodynamics associated with the method \cite{Doye98e} that
makes understanding the physics behind the approach easier, as we shall see in Section \ref{sect:GO_sol}.

Typically, a series of basin-hopping runs of a specified length will be performed starting
from a random geometry. This is advantageous over a single longer run because it can provide a loose gauge of success. 
If all the runs return the same lowest-energy structure one would imagine that the true 
global minimum had been found. It can also often prove useful to perform runs starting from the best 
structures at sizes one above and below, with the lowest-energy atom removed or an atom added, respectively.

The local minimization method that we have found to be most efficient for clusters
is a limited memory BFGS algorithm \cite{Liu89}.
The basin-hopping approach is also found to be more efficient when the configuration is reset to the configuration of the
local minimum after each accepted step \cite{White98a}; this avoids problems with evaporation of atoms
from the cluster since the basin-hopping transformation also reduces the barriers to dissociation. 
In addition to the usual steps, it is advantageous to have occasional angular steps for low-energy surface atoms.
These have a similar aim to the directed mutations introduced by Hartke into his genetic algorithm \cite{Hartke99}; 
they both enable the best arrangement of the surface atoms to be found more rapidly.
For biopolymers other system-specific step types have been introduced to increase efficiency \cite{Derreumaux00a}.

One of the main differences between the basin-hopping and genetic algorithms is that 
only local moves are used in basin-hopping, whereas genetic algorithms employ co-operative 
crossover moves in which a new structure is formed from fragments of two `parent' clusters.
However, recently non-local moves have been introduced into a variant of basin-hopping by rotating
or reflecting a fragment of the structure \cite{Rata00}.

\begin{figure}
\begin{center}
\includegraphics[width=8.5cm]{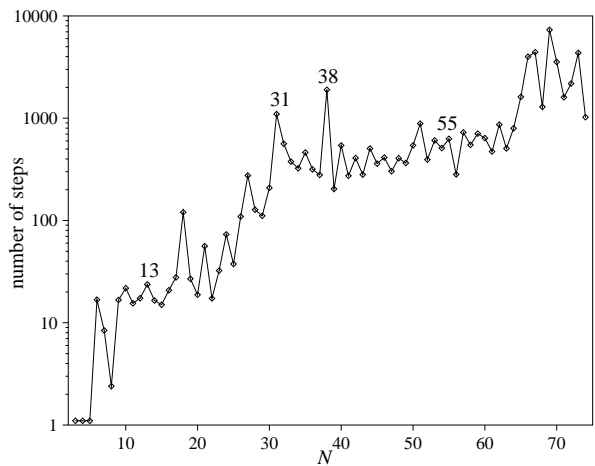}
\end{center}
\vglue -0.5cm
\caption{\label{fig:BH} Mean number of steps to reach the LJ$_N$ (up to $N$=74) 
global minimum from a random starting point with the basin-hopping approach. 
The average is over a hundred runs at $T=0.8\epsilon k^{-1}$.
Reproduced from Ref.\ \cite{WalesS99}.
}
\vglue -0.3cm
\end{figure}

\begin{figure}
\begin{center}
\includegraphics[width=8.5cm]{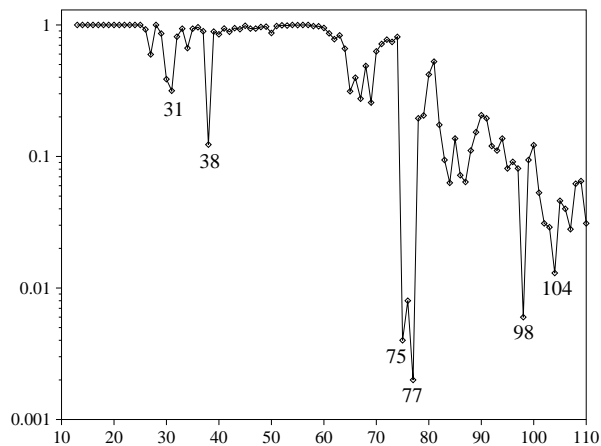}
\end{center}
\vglue -0.5cm
\caption{\label{fig:Leary_MSBH} Observed probability of hitting the global minimum in 
a monotonic sequence basin-hopping run starting from a random starting point, averaged over 1000 runs. 
Reproduced from Ref.\ \cite{Leary00}.
}
\vglue -0.3cm
\end{figure}

Two examples of the performance of basin-hopping algorithms for LJ clusters as a function of 
size are given in Figures \ref{fig:BH} and \ref{fig:Leary_MSBH}. 
The basin-hopping transformation must lead to a considerable speed-up (in terms of steps) 
if the method is to be cost-effective, because the transformation is computationally expensive and requires many evaluations
of the energy and the forces at each step. 
Clearly, one would not want to use the many millions of steps and cycles
that are typically used in molecular dynamics and MC simulations on the original PES.
However, the results in Figure \ref{fig:BH} show that the number of steps required to find the global minimum
is remarkably few, only of the order of hundreds or thousands of steps.
This is even more remarkable when the number of minima on the PES is considered.
In line with theoretical expectations \cite{Still99} the number of minima for small LJ clusters 
increases exponentially with size \cite{Tsai93a}. Extrapolating this trend provides, for example, 
an estimate of $10^{21}$ minima for LJ$_{55}$. 
Therefore, a Levinthal-type paradox \cite{Levinthal}\footnote{A copy of Levinthal's difficult-to-locate 
citation classic (Ref.\ \cite{Levinthal}) can now be found on the web at \\
http://brian.ch.cam.ac.uk/$\sim$mark/levinthal/levinthal.html.} can be formulated 
for locating the global minimum of a cluster: the number of minima of an atomic cluster quickly becomes so 
large that beyond a fairly small size, if these minima were {\it randomly} searched, even at an extraordinarily 
fast rate, it would take an unfeasibly long time to locate the global minimum---so 
how is it possible to find the global minimum?
Yet the basin-hopping algorithm using optimal parameters finds the LJ$_{55}$ global minimum from a
random starting configuration on average within 150 steps \cite{WalesS99}.
The fallacy in Levinthal's paradox has long been known to be the assumption of random searching 
\cite{Zwanzig92,Zwanzig95,Bryngelson95,Doye96c}
however it does emphasize the extremely non-random nature of basin-hopping for LJ$_{55}$---the runs are
extremely biased towards the global minimum.

Figure \ref{fig:BH} shows some interesting variations in the ease of global optimization with cluster size. 
The 38-atom global minimum particularly stands out as being difficult to locate,
suggesting that competition between different structural types makes global optimization
more difficult. Indeed for LJ$_{75}$, the next largest cluster with a non-icosahedral global minimum,
the number of steps required is so large that it was not possible to obtain good enough statistics to 
be included in Figure \ref{fig:BH}.
Subsequent calculations on a super-computer by Leary suggest that the 
mean first passage time is of the order of $10^7$ steps \cite{Leary00}. 
The other six non-icosahedral global minima for $N<150$ are of roughly similar difficulty to locate.
It is also noteworthy that the global minimum for LJ$_{31}$ is relatively difficult to find. In this case,
there is some structural competition between the two types of icosahedral overlayer---it is the first size
at which the overlayer that leads to the next Mackay icosahedron is lowest in energy.

An alternative perspective on these effects can be obtained from Figure \ref{fig:Leary_MSBH},
which shows the probability that a `monotonic sequence' basin-hopping run ends at the global minimum.
In this variation of the basin-hopping algorithm \cite{Leary00} only downhill steps are accepted (i.e. $T$=0)
and the run is stopped after there is no further improvement for a certain number of steps.
For those sizes with non-icosahedral global minima there is a much smaller probability of the
run ending in the global minimum, and so again global optimization is more difficult.
In these cases the majority of runs end at low-energy icosahedral structures.
Those examples at larger sizes are again more than an order of magnitude more difficult than LJ$_{38}$,
but interestingly the Marks decahedra at $N$=102--104 are somewhat more easy to locate 
than those at $N$=75--77.

These results enable us to comment on the use of LJ clusters as a test system for GO methods.
They show that the icosahedral global minima are relatively easy to locate, and 
in these cases optimization only starts to become more difficult as $N$ approaches 100 (Figure \ref{fig:Leary_MSBH}). 
As for the non-icosahedral global minima, the number of unbiased GO methods that have found 
the LJ$_{38}$ global minimum is now quite large 
\cite{WalesD97,Pillardy,Deaven96,Hartke99,Rata00,Leary00,Niesse96a,Michaelian98,Pappu98,Pillardy99,Faken99,Locatelli,Neirotti00}
but those that can find the LJ$_{75}$ global minimum is still small \cite{WalesD97,Hartke99,Rata00,Leary00,Locatelli}.
Therefore, a good test for a GO method is to attempt to find all the global minima up to $N$=110.
Any GO method `worth its salt' for clusters should be able find all the icosahedral global minima and 
the truncated octahedron at $N$=38 . Success for the other non-icosahedral global minima would 
indicate that the method has particular promise.
However, far too many GO algorithms have only been tested on cluster 
sizes where global optimization is relatively trivial.

The weakness of LJ clusters as a test system is that they have a relatively uniform 
structural behaviour.  Morse clusters could provide a much more varied test system, 
as Figure \ref{fig:Mphased} illustrates.
A suitable test would be to aim to find all the global minima at $\rho$=3, 6, 10 and 14 up to $N$=80, 
as putative global minima have been tabulated for this size and parameter range \cite{Doye95c,Doye97d}.
One would generally expect the difficulty of global optimization to increase with $\rho$ because the number of minima
increases \cite{Doye96b,Miller99a} and the energy landscape becomes more rough \cite{Miller99a,Miller99b}.
The system also provides many examples of structural competition, particularly for the short-ranged potentials
where decahedral and close-packed clusters can have similar energies.
A number of studies have begun to use Morse clusters as a test system \cite{Roberts00,Xu00}.

\section{Multiple-Funnel Energy Landscapes}

The aim of this section is to 
provide a physical perspective that can help us understand why the 
global optimization of a system is easy or difficult, for example, to explain the size-dependence of 
Figures \ref{fig:BH} and \ref{fig:Leary_MSBH}.
As mentioned in Section 3,
an equivalent of Levinthal's paradox, which was originally
formulated to capture the difficulty of a protein folding to its native state, can be applied 
to a cluster locating its global minimum.
The flaw in that paradox is its assumption that conformations will be sampled randomly, 
i.e.\ all configurations are equally likely, because we know that in 
an equilibrium physical sampling of the conformation space, say in the canonical ensemble, each point will be
sampled with a probability proportional to the Boltzmann weight, $\exp(-E({\bf x})/kT)$.
Thus, the Boltzmann factor favours low-energy conformations.
Therefore, we can begin to see the vital role played by the potential energy surface.
This role extends beyond purely thermodynamic considerations to the dynamics:
does the topography and connectivity of the PES naturally lead the system towards or away from the global minimum?

The Levinthal assumption of random sampling is equivalent to assuming that the energy landscape has the
topography of a perfectly flat putting green with no thermodynamical or dynamical biases towards the global minimum
at the bottom of the `hole'. Similarly, the NP-hard character of the global optimization of atomic clusters \cite{Wille},
which results in part from the exponential increase in the search space with size, considers 
a general case where no assumptions about the topography of the PES can be made.
In the protein folding community, after the fallacy in the Levinthal paradox was recognized,
attention focussed on the more important question of how does the topography of the PES
differ for those polypeptides that are able to find their native states from those that cannot \cite{Bryngelson95}.
Here, I address similar questions for the global optimization of clusters.

\begin{figure}
\includegraphics[width=12cm]{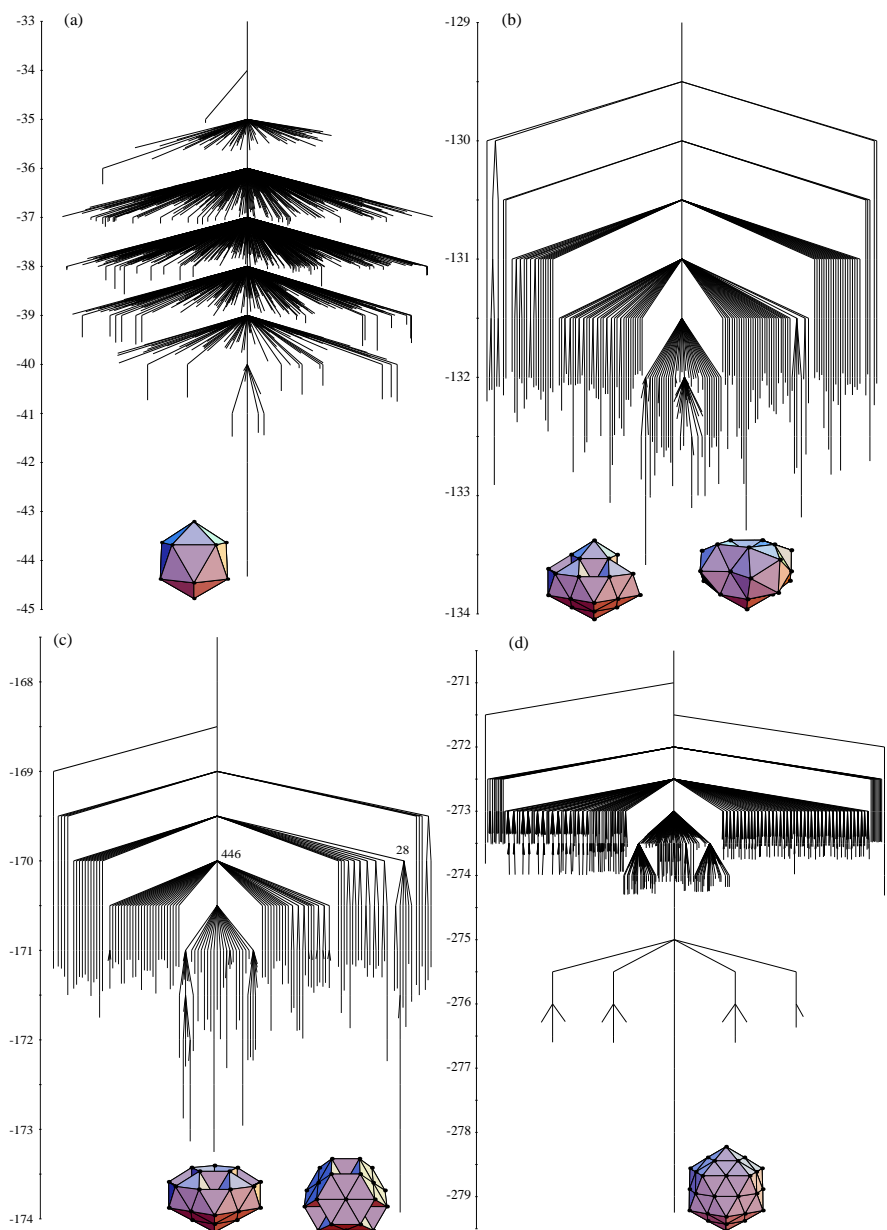}
\vglue -0.5cm
\caption{\label{fig:LJtree} 
Disconnectivity graphs for
(a) LJ$_{13}$, (b) LJ$_{31}$, (c) LJ$_{38}$,
(d) LJ$_{55}$, (e) LJ$_{75}$ and (f) LJ$_{102}$.
In (a) all the minima are represented.
In the other parts only the branches leading to the
(b) 200, (c) 150, (d) 900, (e) 250 and (f) 200 lowest-energy minima are shown.
The numbers adjacent to the nodes indicate the number of minima the nodes represent.
Pictures of the global minimum, and sometimes the second lowest-energy minimum,
are adjacent to the corresponding branch. 
The units of the energy axes are $\epsilon$. 
}
\end{figure}

\addtocounter{figure}{-1}
\begin{figure}
\includegraphics[width=12cm]{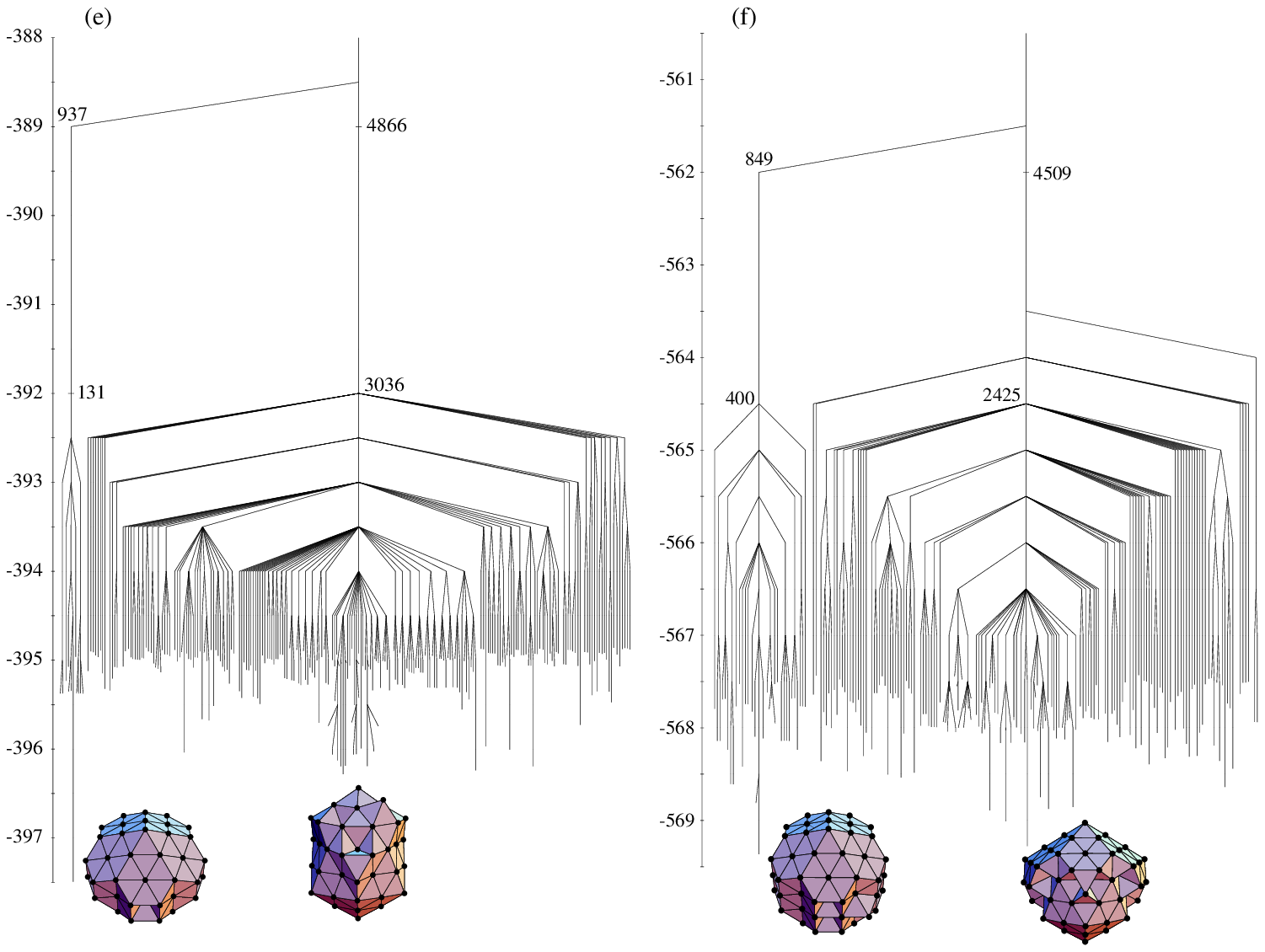}
\vglue -0.5cm
\caption{cont.}
\end{figure}

One of the topographical features of the energy landscape that the protein folding community 
has found to be common is, what has been termed, a `funnel' \cite{Bryngelson95,Leopold}. 
By this they mean a region of configuration space that can be described in terms of a set 
of downhill pathways that converge on a single low-energy structure or a set of 
closely-related low-energy structures. As its name suggest a protein PES with a single funnel 
converging on the native state will be a good folder because the topography helps in
guiding the protein towards that native state.

If these ideas are to be useful, one needs a way of depicting the physically-relevant aspects
of the topography of a complicated $3N$-dimensional energy landscape.
One technique that has proven to be helpful in characterizing the PESs of 
proteins \cite{Becker97,Levy98a,Miller99c,WalesDMMW00} and clusters \cite{WalesDMMW00,WalesMW98,Doye99f} is the 
disconnectivity graph. 
This graph provides a representation of the connectivity of the multi-dimensional 
energy landscape and by depicting the effective barriers between minima 
it is especially useful in the interpretation of dynamics.

To construct a disconnectivity graph, at a series of energy levels the minima on the PES 
are divided into sets which are connected by paths that never exceed that
energy level. In the graph each set is represented by a node at the appropriate energy 
and lines connect a node to the sets at higher and lower energy which contain the
minima corresponding to the original node. 
A line always ends at the energy of the minimum it represents. 

Disconnectivity graphs can be understood by analogy to the effects of the water level in a geographical landscape. 
The number of nodes in a graph at a given energy is equivalent to the number of distinct seas 
and lakes for a given water level.
Inspection of actual disconnectivity graphs,
e.g.\ Figure \ref{fig:LJtree}, also helps to clarify these ideas. 
At sufficiently high energy, all minima are mutually accessible and so there is only one node, 
but as the energy decreases sets of minima become disconnected from each other and the graph splits, 
until at sufficiently low energy, again there is only one node left, that of the global minimum.
The pattern of the graph can reveal particularly interesting information about the 
PES topography. For a PES with a single funnel there is a single dominant stem with the other minima
branching directly off it as the energy is decreased. By contrast for `multiple-funnel' PESs the 
graph is expected to split at high energy into two or more major stems.

In Figure \ref{fig:LJtree} disconnectivity graphs for a selection of LJ clusters are presented,
in particular some of those clusters that Figures \ref{fig:BH} and \ref{fig:Leary_MSBH} indicated 
are more difficult to optimize. In the graph for LJ$_{13}$ all the minima in our near-exhaustive sample 
are represented. The graph shows the form for an ideal single-funnel PES, because the 
icosahedral global minimum is particularly low in energy, and dominates the energy landscape. 
The PES has a remarkable connectivity: 911 distinct transition states are connected to the global minimum
and all minima are within three rearrangements of the global minimum.
The disconnectivity graph for LJ$_{55}$, another `magic number' LJ cluster, also has a single-funnel. 
Unlike for LJ$_{13}$, we are unable to represent all the minima that we found on the graph, 
so instead we concentrate on the lower-energy minima in our sample. 
Indeed, the two bands of minima in the graph represent Mackay icosahedra with one or
two defects. The graph only reveals the bottom of a funnel which extends up into the 
liquid-like minima \cite{Doye99f}.

\begin{figure}
\begin{center}
\includegraphics[width=8cm]{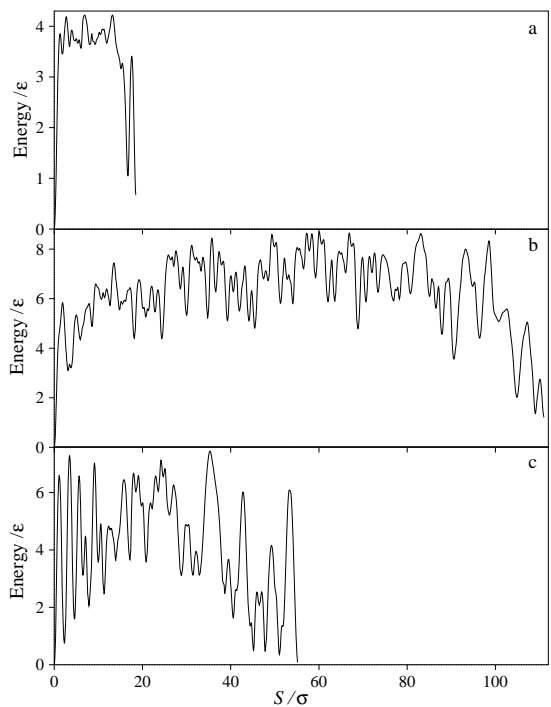}
\end{center}
\vglue -0.5cm
\caption{\label{fig:LJ_epath} The lowest-energy path from the global minimum to the second 
lowest-energy minimum for (a) LJ$_{38}$, (b) LJ$_{75}$ and (c) LJ$_{102}$. In each
case the zero of energy corresponds to the energy of the global minimum.
}
\vglue -0.3cm
\end{figure}

In contrast to these two clusters, the bottom of the LJ$_{31}$ PES is much flatter, and there are significant
barriers between the low-energy minima, in particular between the two lowest-energy minima, which are 
icosahedral structures, but with different types of surface overlayer.
These effects of structural competition are found 
in more extreme form in the graphs of those clusters with non-icosahedral global minima. The graphs of
LJ$_{38}$, LJ$_{75}$ and LJ$_{102}$ split at high energy into stems associated with icosahedral
and fcc or decahedral structures, and so these energy landscapes have two major funnels. 
This splitting is most 
dramatic for LJ$_{75}$ and LJ$_{102}$, where the barrier between the two funnels is much greater
than between any of the other sets of minima.
The barriers between the two lowest-energy minima are $4.22\epsilon$ and $3.54\epsilon$ for LJ$_{38}$,
$8.69\epsilon$ and $7.48\epsilon$ for LJ$_{75}$ and $7.44\epsilon$ and $7.36\epsilon$ for LJ$_{102}$.
The corresponding lowest-barrier paths are represented in Figure \ref{fig:LJ_epath}.
These pathways pass over many transition states---13, 65 and 30 for 
LJ$_{38}$, LJ$_{75}$ and LJ$_{102}$, respectively. 
There are many other pathways connecting the funnels, 
but they are either longer or involve higher effective barriers.
For LJ$_{38}$ the pathway passes through disordered liquid-like minima. 
However, for the two larger clusters all the minima along the 
pathways are ordered and the main structural changes are achieved by rearrangements that 
involve cooperative twists around the five-fold axis of the decahedron---the
conservation of this axis throughout the structural transformation has also been observed in simulations of the 
decahedral to icosahedral transition in gold clusters \cite{Cleveland98}. 
For LJ$_{75}$ the Marks decahedron is oblate whilst the low-energy icosahedral minima are
prolate, so the pathway involves a greater amount of reorganization of the surface layer
either side of these cooperative transitions than for LJ$_{102}$. For the laterr cluster the decahedral
and icosahedral structures have fairly similar shapes; hence the shorter pathway for LJ$_{102}$ 
(Figure \ref{fig:LJ_epath}), even though it is larger.

\begin{figure}
\includegraphics[width=12cm]{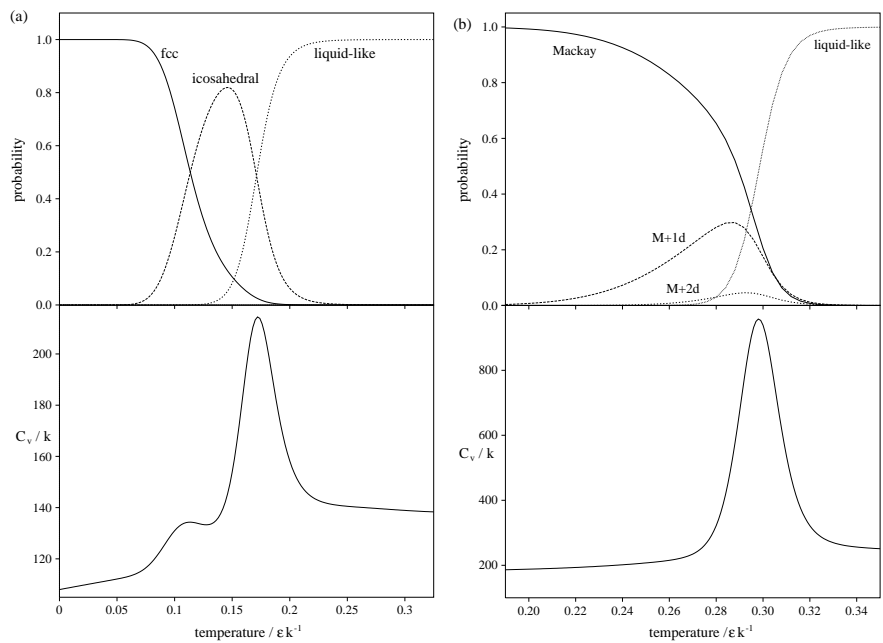}
\vglue -0.3cm
\caption{\label{fig:LJ_therm} Equilibrium thermodynamic properties of (a) LJ$_{38}$ and (b) LJ$_{55}$
in the canonical ensemble.
Both the probability of the cluster being in the labelled regions of configuration space
and the heat capacity, $C_v$ are depicted. The label M+$n$d stands for a Mackay icosahedron
with $n$ surface defects.
}
\vglue -0.3cm
\end{figure}

As expected from the general dominance of icosahedral structures in this size range, there 
are many more low-energy icosahedral minima than low-energy decahedral or fcc minima for
these three clusters. There are many low-energy arrangements of 
the incomplete surface layer of the icosahedral structures, whereas the
decahedral or fcc structures have especially compact structures and so any alteration 
in structure leads to a significant increase in energy.
The number of minima in our samples that lie within the respective funnels is marked
on Figure \ref{fig:LJtree} and is indicative of the greater width of the icosahedral funnels.

From characterizing the energy landscapes of these clusters, the physical origins of 
some of the differences in difficulty for GO algorithms should be becoming 
apparent. The single funnels of LJ$_{13}$ and LJ$_{55}$ make the global minimum particularly 
accessible, and the system is strongly directed towards the global minimum on relaxation
down the PES. For LJ$_{31}$, once the system has reached a low-energy structure, 
the flatness and the barriers at the bottom of the PES 
mean that further optimization is relatively slow compared to LJ$_{13}$ and LJ$_{55}$.
For the clusters with non-icosahedral global minima, the icosahedral funnel is much more 
accessible because of its greater width. Furthermore, after entering the icosahedral funnel, 
subsequent escape into the fcc or decahedral funnel is likely to be very slow because of the large
barriers that need to be overcome. The icosahedral funnel acts as a kinetic trap 
hindering global optimization.
These effects will come out even more clearly in the next two sections as we look
at the thermodynamics and dynamics associated with these clusters.
From the disconnectivity graphs one would expect trapping to be a significantly
greater hindrance to global optimization for LJ$_{75}$ and LJ$_{102}$ than
for LJ$_{38}$, and the longer interfunnel pathway for LJ$_{75}$ provides a possible 
explanation of why LJ$_{102}$ seems to be somewhat less difficult to optimize than
LJ$_{75}$.

\subsection{Thermodynamics}
\label{sect:therm}

\begin{figure}
\begin{center}
\includegraphics[width=8cm]{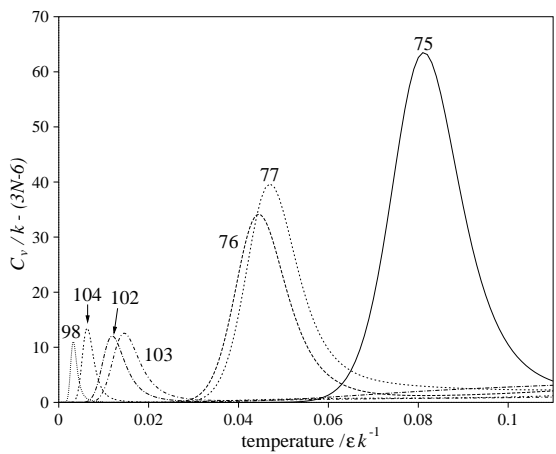}
\end{center}
\vglue -0.5cm
\caption{\label{fig:LJss} 
Canonical heat capacity peaks associated with the structural transitions from the
global minimum to icosahedral structures for the LJ clusters
with non-icosahedral global minima. The sizes are as labelled.
}
\vglue -0.3cm
\end{figure}

The typical thermodynamic properties of a cluster are illustrated in Figure \ref{fig:LJ_therm}b
for LJ$_{55}$. The heat capacity peak is associated with a melting transition that is the finite-size
analogue of a first-order phase transition \cite{Labastie} and up to melting the structure is based on 
the global minimum, perhaps with some surface defects. The thermodynamics 
of the clusters with non-icosahedral structures are significantly different. Now, as well as the
melting transition there is a further transition associated with a transition
from the global minimum to the icosahedral structures that gives rise to a second lower-temperature 
peak in the heat capacity (Figure \ref{fig:LJ_therm}a and \ref{fig:LJss}).
For LJ$_{38}$ the transition occurs fairly close to melting,\footnote{The 
thermodynamic properties illustrated in Figures \ref{fig:LJ_therm} and \ref{fig:LJss} have 
been calculated using a method where the thermodynamic properties of the individual minima are 
summed \cite{WalesDMMW00,Doye95a}.
However, recent simulations using parallel tempering indicate that for LJ$_{38}$ the two transitions are
slightly closer than indicated by Figure \ref{fig:LJ_therm} and so the fcc to icosahedral transition
results only in a shoulder in the heat capacity curve \cite{Neirotti00}.}
however as the size increases the transition temperature generally decreases \cite{Doye00d}. 

These solid-solid transitions have a number of implications for global optimization. 
Firstly, on cooling from the melt it is thermodynamically more favourable for the cluster to 
enter the icosahedral funnel than that associated with the global minimum. Secondly, the transitions,
particularly those for the larger clusters, lie below the `glass transition' temperature where
the cluster is effectively trapped in the current local minimum. This presents a nightmare
scenario for simulated annealing. On cooling the cluster would first enter the icosahedral funnel,
where it would then become trapped, even when the global minimum becomes thermodynamically more
stable, because of the large free energy barriers (relative to $kT$) for escape from this funnel \cite{Doye99c}.

In the protein folding literature, good folders have been shown to have a large value of
the ratio of folding temperature to the glass transition temperature, $T_f/T_g$, because
this ensures the kinetic accessibility of the native state of the protein at temperatures
where it is thermodynamically most favoured \cite{Bryngelson95}. 
By contrast, these clusters have effective $T_f/T_g$ values less than one 
and are archetypal `bad folders'.

\subsection{Dynamics}

It is impractical to examine the interfunnel dynamics by standard molecular dynamics simulations
because of the extremely long time scales involved. However, it is possible to calculate the 
rate of interfunnel passage by applying a master equation approach\footnote{In the master equation approach
the occupation probabilities of all the minima can be followed as a function of time, given a set 
of rate constants between adjacent minima. These rate constants can be approximated using standard
rate theories \cite{Forst}.} 
to the large samples of minima and transition states used to construct the disconnectivity 
graphs \cite{Miller99b,WalesDMMW00}.
The rate constants for LJ$_{38}$ have been computed and show
that the interfunnel dynamics obeys an Arrhenius law (i.e. the rate is proportional to $\exp(-E_a/kT)$, 
where $E_a$ is the activation energy) well with the activation energies
corresponding to the barriers associated with the lowest-energy pathway 
between the two funnels \cite{Miller99b,WalesDMMW00}. For LJ$_{38}$ this gives a value of
43\,s$^{-1}$ for the interfunnel rate constant at the centre of the fcc to icosahedral 
transition (using parameters appropriate for Ar). As exected, this is beyond the time scales 
accessible to molecular dynamics simulations.
The equivalent rate constants for the larger clusters are much slower because of the larger
activation energies and the lower transition temperatures.

\begin{figure}
\begin{center}
\includegraphics[width=7cm]{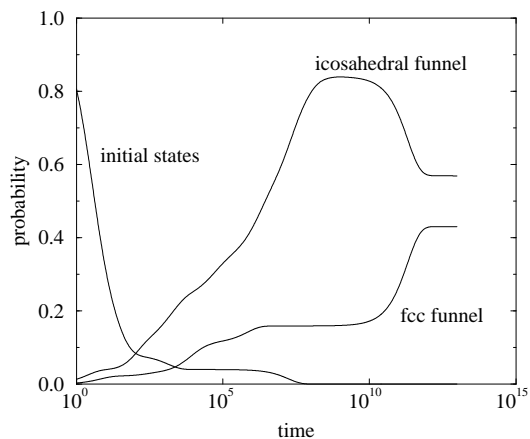}
\end{center}
\vglue -0.8cm
\caption{\label{fig:LJ38dyn} 
Relaxation of LJ$_{38}$ from high-energy minima showing the fast and slow contributions to the 
final probability of the fcc funnel. The time is in units of $(m\sigma^2/\epsilon)^{1/2}$.
}
\vglue -0.3cm
\end{figure}

Figure \ref{fig:LJ38dyn} illustrates for LJ$_{38}$ the dynamics of relaxation from high-energy states.
The initial relaxation is relatively rapid and the majority of the population enters the 
icosahedral funnel (the approximations in this calculation actually lead to an overestimation
of the probability of initially entering the fcc funnel). These processes are separated by a couple of decades in
time from the subsequent equilibration between the two funnels.

The combined effects of the thermodynamics and dynamics can be illustrated by some simulated annealing results.
For LJ$_{55}$ the probability of reaching the global minimum in annealing simulations of $10^6$ and $10^7$ MC cycles
is 29\% and 94\%. The equivalent values for LJ$_{38}$ are 0\% and 2\%, and for LJ$_{75}$ the annealing 
simulations were never able to locate the global minimum.

\subsection{Optimization solutions}
\label{sect:GO_sol}

The previous sections illustrate the difficulty of finding the global minimum of the LJ clusters
with non-icosahedral global minima if the natural thermodynamics and dynamics of the system are followed.
However, optimization approaches do not have to be restricted to this behaviour. 
For example, as we mentioned in Section 3
the basin-hopping transformation 
accelerates the dynamics, allowing hops directly between basins.
However, the transformation only reduces the interfunnel energy barriers
by $0.68\epsilon$ for LJ$_{38}$, $0.86\epsilon$ for LJ$_{75}$ and $0.89\epsilon$ for LJ$_{102}$.
Therefore, multiple-funnels are still potentially problematic.

\begin{figure}
\includegraphics[width=12cm]{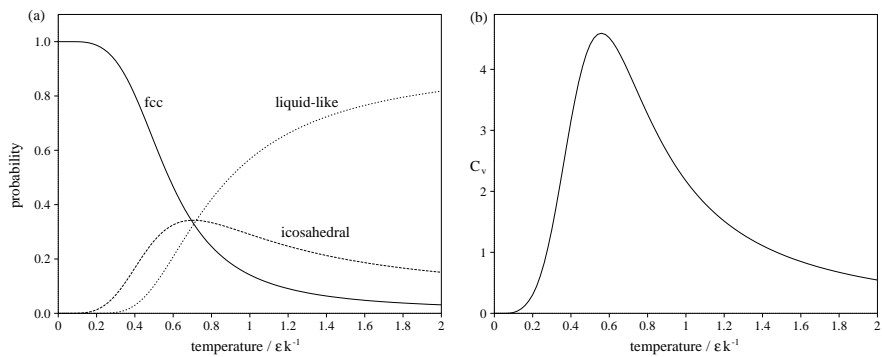}
\vglue -0.3cm
\caption{\label{fig:LJ38_gmin_therm} Equilibrium thermodynamic properties of LJ$_{38}$ on
the transformed PES in the canonical ensemble.
(a) The probability of the cluster being in the fcc, icosahedral and liquid-like regions of 
configuration space.
(b) The configurational heat capacity, $C_v$.
}
\vglue -0.3cm
\end{figure}

\begin{figure}
\includegraphics[width=10cm]{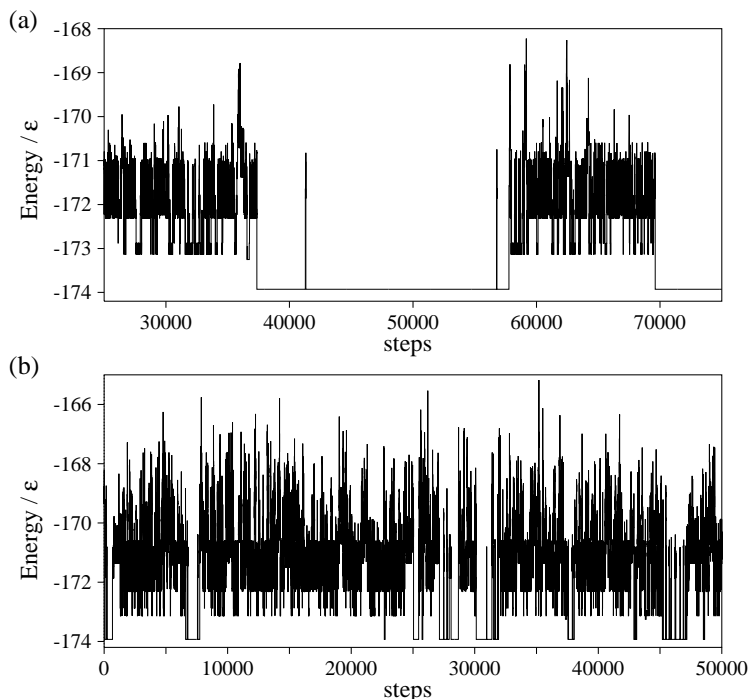}
\vglue -0.3cm
\caption{\label{fig:gmin_dynamics} $\tilde{E}$ as a function of the number of steps
in basin-hopping runs for LJ$_{38}$ at (a) $T$=0.4$\epsilon k^{-1}$ and 
(b) $T$=0.9$\epsilon k^{-1}$.
}
\vglue -0.3cm
\end{figure}

Figure \ref{fig:LJ38_gmin_therm} shows that the thermodynamic properties of LJ$_{38}$
are dramatically changed by the transformation. The transitions have been significantly broadened. 
There is now only a single heat capacity peak and a broad temperature range where all states
are populated. In particular, the global minimum now has a significant probability of occupation
at temperatures where the free energy barriers between the funnels can be surmounted. This
effect is illustrated by the basin-hopping simulations in Figure \ref{fig:gmin_dynamics}. 
At low temperature the system is localized in one of the funnels with transitions between 
the two funnels only occurring rarely. However, at higher temperatures transitions
occur much more frequently \cite{Doye98e}. As we noted earlier the rate of interfunnel passage is proportional to 
$\exp(-E_a/kT)$. Although the transformation only reduces $E_a$ by a small amount, the temperature
for which the occupation probability for the global minimum still has a significant value is now over
ten times larger. Hence, the increased interfunnel rates.

The transformation has a second kinetic effect: it increases the width of the funnel of 
the global minimum. On the original LJ$_{38}$ PES only 2\% of the long annealing runs entered the icosahedral
funnel, whereas 13\% of Leary's downhill basin-hopping runs ended at the global minimum 
(Figure \ref{fig:Leary_MSBH}).
On relaxation down the energy landscape the system is much more likely to enter the fcc 
funnel on the transformed PES, thus making global optimization easier.

These changes to the thermodynamics and dynamics also reduce the difficulty of global optimization for the larger
non-icosahedral clusters, making it possible, if still very difficult, to reach the global minimum. 
The results of Leary indicate that for these clusters the increased accessibility of the funnel of the
global minimum is the more important effect of the transformation. He found that it is more efficient 
to restart a run when it gets stuck in a funnel (and hope that it enters the funnel of the global minimum next time) 
than to wait for the cluster to escape from that funnel \cite{Leary00}.

We can understand why the PES transformation so dramatically changes the thermodynamics of the system, 
by examining $p_i$, the occupation probability of a minimum $i$.
For the untransformed PES within the harmonic approximation
$p_i\propto \exp(-\beta E_i)/\overline{\nu_i}^{3N-6}$, where $E_i$ is the potential energy of minimum 
$i$ and $\overline{\nu_i}$ is the geometric mean vibrational frequency.
For the transformed PES $p_i\propto A_i \exp(-\beta E_i)$, where $A_i$ is the hyperarea
of the basin of attraction of minimum $i$. The differences between these expressions,
the vibrational frequency and hyperarea terms, have opposite effects on the thermodynamics. 
The higher-energy minima are generally less rigid
and so the vibrational term entropically stabilizes the icosahedra and, even more so, the liquid-like state,
pushing the transitions down to lower temperature and sharpening them.
By contrast the hyperarea of the minima decreases with increasing potential energy, thus stabilizing the 
lower-energy states and broadening the thermodynamics.

\section{Conclusions}

In this chapter I hope to have shown how physical insight can play an important role in
understanding the behaviour of global optimization algorithms
and hope that these insights can help provide a firmer physical (rather than just empirical or intuitive)
foundation for the design of new improved algorithms. I also hope that it will encourage some
to think about the physical aspects of other types of GO problems.

In particular I have highlighted some examples where a multiple-funnel energy landscape
strongly hinders global optimization. Such challenging cases are likely to be a common feature
for any cluster system where there is not a single strongly dominant structural type. More generally,
multiple funnels probably represent the most difficult problem for this class of GO problems,
in which the lowest-energy configuration of a system is sought.
For example, for this reason, the main criterion in the design of polypeptides that fold well is the 
avoidance of multiple funnels by optimizing the energy gap between the native state and 
competing low-energy structures \cite{Goldstein92}.

I have also shown why the basin-hopping transformation of the energy landscape
makes global optimization for these multiple-funnel cases easier. The method's
success results from a broadening of the thermodynamics, so that
the occupation probability of the global minimum is significant at temperatures where 
the interfunnel free energy barriers can be surmounted. 
This idea should act as a design principle in the development of any new GO algorithms
that hope to overcome the challenge of multiple funnels. 

Although basin-hopping is successful for the LJ multiple-funnel examples,
further improvements in efficiency are required before one can hope to succeed 
for similar cases at larger size or for potential energy functions that
are significantly more computationally intensive to evaluate.
A number of avenues by which this may be achieved suggest themselves.
Firstly, the basin-hopping algorithm searches the transformed potential energy
surface using simple MC. However, there are a whole raft of methods that
have been developed in order to speed up the rate of rare events in
simulations on the untransformed PES, 
which could potentially be applied to the transformed PES. 
These include parallel tempering \cite{Marinari92}, jump-walking \cite{Frantz90}, 
and the use of non-Boltzmann ensembles, such as Tsallis statistics \cite{Andricoaei96}.

Secondly, the basin-hopping transformation can be combined with other PES transformations,
of which there have been many suggestions \cite{Schelstrate99}, to get a double transformation 
(and hopefully a greater simplification) of the PES \cite{Doye00c}. 
The potential problem with this type of approach is that as
well as smoothing the PES, most transformations also change the relative stabilities of
different structures. This can work in one's favour if the transformation stabilizes the
global minimum. For example, a recent transformation proposed by Locatelli and Schoen, which favours compact 
spherical clusters, stabilizes the non-icosahedral LJ global minima \cite{Locatelli}; 
for the 38-atom cluster the PES, when sufficiently deformed, has a single-funnel topography 
with the truncated octahedron at its bottom \cite{Doye00c}. 
But just as often a transformation will destabilize the global minimum---this is why
the performance of many PES-transformation GO methods is erratic, perhaps solving some 
`hard' instances while failing for `easier' examples. However, 
the proposed approach could be run alongside standard basin-hopping runs, 
and so would perhaps succeed in instances where the standard 
algorithm struggles.

Thirdly, Hartke recently proposed a modification to the genetic algorithm approach,
in which a diversity of structures is maintained in the population \cite{Hartke99}. 
This leads to significant increases in efficiency for the LJ clusters with 
multiple-funnels, because it prevents the whole population being confined (and trapped)
within the icosahedral funnel. Such an approach could also potentially increase the efficiency
of other algorithms. For example, a diversity of structures could be maintained between a 
set of parallel basin-hopping runs.

\begin{acknowledgments}
The author is the Sir Alan Wilson Research Fellow at Emmanuel College, Cambridge.
I would like to acknowledge the role played by David Wales and Mark Miller in
this research program and to thank Bob Leary for providing the data for Figure \ref{fig:Leary_MSBH}.
\end{acknowledgments}

\end{document}